## Strain doping: Reversible single axis control of a complex oxide lattice via helium implantation

Hangwen Guo[1,2,3], Shuai Dong[1,2,4], Philip D. Rack[5,6], John D. Budai[1], Christianne Beekman[1], Zheng Gai[5], Wolter Siemons[1], C.M. Gonzalez[6], R. Timilsina[6], Anthony T. Wong[1,6], Andreas Herklotz[1], Paul C. Snijders[1,2], Elbio Dagotto[1,2], Thomas Z. Ward[1*]

[1] Materials Science and Technology Division, Oak Ridge National Laboratory, Oak Ridge, TN, 37831, USA
[2] Department of Physics & Astronomy, University of Tennessee, Knoxville, TN 37996, USA
[3] Department of Physics & Astronomy, Louisiana State University, Baton Rouge, LA 70803, USA
[4] Department of Physics, Southeast University, Nanjing, 211189, China
[5] Center for Nanophase Materials Sciences, Oak Ridge National Laboratory, Oak Ridge, TN, 37831, USA
[6] Materials Science & Engineering, University of Tennessee, Knoxville, TN 37996, USA

We report on the use of helium ion implantation to independently control the out-of-plane lattice constant in epitaxial $La_{0.7}Sr_{0.3}MnO_3$ thin films without changing the in-plane lattice constants. The process is reversible by a vacuum anneal. Resistance and magnetization measurements show that even a small increase in the out-of-plane lattice constant of less than 1% can shift the metal-insulator transition and Curie temperatures by more than 100 °C. Unlike conventional epitaxy-based strain tuning methods which are constrained not only by the Poisson effect but by the limited set of available substrates, the present study shows that strain can be independently and continuously controlled along a single axis. This permits novel control over orbital populations through Jahn-Teller effects, as shown by Monte Carlo simulations on a double-exchange model. The ability to reversibly control a single lattice parameter substantially broadens the phase space for experimental exploration of predictive models and leads to new possibilities for control over materials' functional properties.

The crystal lattice is one of the most accessible degrees of freedom in materials. In complex oxides, effective control over lattice parameters not only facilitates the understanding of multiple interactions in strongly correlated systems, but also creates new phases and emergent functionalities [1–4]. Lattice engineering has played an extremely important role in attempts to design strongly correlated systems and has led to many important discoveries [1,5–8]. Control over lattice strain in films using different substrates [9] has revealed enhanced ferroelectricity [10] and superconductivity [11], as well as induced superconductivity in otherwise non-superconducting compounds [12]. However, the basic nature of the broken translational symmetry in the crystal lattice also entails a rigidity against arbitrary control [13]. There is so far no experimental technique that allows one to alter the lattice parameter solely along a *single* crystal axis, i.e. with an effective Poisson's ratio of zero. For strain engineering in systems with a non-zero Poisson ratio, the lattice constant, and hence the electronic structure, necessarily change in all three directions, clouding the cause-effect relations between single degrees of freedom and order parameters.

We demonstrate an approach using helium implantation to effectively "strain dope" the lattice along a single axis of a $La_{0.7}Sr_{0.3}MnO_3$ (LSMO) film that is epitaxially lattice-locked to a substrate. The out-of-plane(*c*-axis) lattice constant can be modified independently of the in-plane lattice constants. The *c*-axis strain can be continuously manipulated, and is thus not restricted by the limited collection substrates that dictate conventional epitaxial strain engineering. The change in materials' properties, while reversible via a high temperature anneal, is persistent even well above room temperature. No continuous external actuation is required as with transient pressure-induced [14] or piezo-induced [15] strain states. Monte Carlo simulations on a double-exchange model reveal that such a shift in the out-of-plane lattice degree of freedom directly modifies the orbital occupancies through Jahn-Teller coupling, which in turn drives changes in phase transition temperatures and closely match our experimental observations. Our approach opens a persistent, previously inaccessible, and continuously tunable phase space to manipulate complex correlated behavior by tuning lattice strain along a single axis and is expected to be applicable in any epitaxially locked thin film or near-surface bulk crystal.

We select LSMO as a model system because of its wide usage, well-known phase diagram, and the existence of well-tested computational models that include the basic interactions and produce qualitative agreement with the experimental phase diagram. LSMO thin films of 20 nm thickness are grown epitaxially on $SrTiO_3(001)$(STO) substrates by pulsed laser deposition [16]. A gold buffer layer is then deposited on the film surface and 4 keV helium ions are injected into the grounded sample [17–29]. The advantages of utilizing helium atoms lie in the fact that helium's stopping power in the lattice comes almost entirely from non-nuclear interactions which minimizes damage to the film structure [29]. Moreover, helium's nobility assures that no extra electrons or holes are doped into the films as with hydrogen doping [30]. The gold buffer layer attenuates the energy and dose of helium ions reaching the film which further reduces the danger of defect generation while removing the impact of surface sputtering from the oxide film. Previous studies without buffer layers show that the doses and energies used in this study are well below the threshold for defect formation of $1x10^{16}$ ions/cm$^2$ at 7 keV [29,31]

Figure 1a shows the *θ-2θ* x-ray diffraction(XRD) scans through the $(002)_{pc}$ peak of the LSMO films before and after helium implantation, where *pc* indicates pseudocubic indices. In all cases, Laue fringes confirm excellent film uniformity. The as-grown LSMO film is found to be tensile strained and epitaxial to the STO substrate with *a=b=*3.905 Å. Tensile strain reduces the out-of-plane, *c*-axis, through the Poisson effect, to 3.8406 Å which gives a Poisson ratio of 0.38 [17], in agreement with previous observations [32]. Upon increasing the He dose, the position of the LSMO$(002)_{pc}$ peak shifts toward the STO$(002)_{pc}$ peak demonstrating an increase in the *c*-axis parameter from the strained tetragonal toward an artificially large cubic state without noticeable loss of peak intensity or addition of spurious phases. The c-axis values were calculated from the XRD line scans using the kinematic fitting method [33]. Figure 1b summarizes the *c*-axis lengths as a function of He dose and reveals a continuous tunability without sharp jumps which would be indicative of strain induced crystal phase reorientation. The percentage increases in the *c*-axis relative to the as-grown state are found to be 0.37%, 0.62%, and 0.94%. Initial implantation modelling suggests that a $2\times10^{15}$ helium dose results in 1 helium atom per a volume of ~$(8.6 \times 8.6 \times 8.6)$ unit cells [17]. Even after the largest He ion dose the film's in-plane lattice is still epitaxially locked to the STO substrate, fully strained, and shows no evidence of in-plane shear-strain-induced twinning [34](Fig. 1c). The strain induced by the He ion thus dissipates along the out-of-plane direction due to the one-dimensional relaxation potential offered by the free surface while the in-plane directions are epitaxially locked by the constraining "infinite" crystal along those directions. Unlike existing strain tuning methods that are unable to manipulate a single crystal axis, these results establish that strain can be independently and continuously controlled along a single axis using He implantation [14,15].

The complex nature of interactions in LSMO means that even small changes to a single lattice parameter can lead to strong modifications of electronic and magnetic properties [7,20,35,36]. The temperature-dependent resistivity is presented in Figure 2a for different He implantation doses. The undosed, as-grown LSMO film exhibits a metal-to-metal transition temperature of 362 K which is typical for this system [32]. As the *c*-axis is expanded, the high temperature metallic phase begins to lose metallicity for an expansion of 0.37%, and becomes insulating for 0.62% and 0.94% expansions. The temperature at which this phase transition occurs is thus strongly tied to the magnitude of the *c*-axis lattice constant, where increasing the *c*-axis leads to a reduction in transition temperature. The magnetoresistance(MR) also shows a high degree of tunability(Fig. 2b). Here, the room temperature MR increases from 25% in the as-grown sample to 90% in the 0.62% expanded sample. The 0.94% expanded sample presents the largest MR value while maintaining a high response across a large temperature range. This is particularly attractive to applications as both MR response and active temperature range can be tuned via a simple post-process helium implantation. Magnetization measurements show similar tunability(Fig. 2c).

As helium is a noble gas and does not form stable compounds, and since we see no evidence of lattice defect formation in our X-ray diffraction data, we conclude that the helium atoms are located interstitially within the lattice. This raises the issue of the thermal stability of the implanted He ions. Figure 3a shows a residual gas analysis of a heavily dosed LSMO film during temperature sweeps between 20 °C to 1250 °C conducted in ultrahigh vacuum. In the first cycle, we observe that helium begins to leave the lattice at 250 °C and reaches a peak loss rate at ~800 °C. A subsequent temperature sweep shows no helium loss. Temperature-dependent resistivity after repeated 3 hour, 0.1 mTorr vacuum anneals with increasing temperature shows that the loss of helium induced by the vacuum annealing process allows the resistive behavior to return toward the as-grown state(Fig. 3b). This also suggests that oxygen stoichiometry is not a factor in the observed *c*-axis expansion and behavioral changes in the helium implanted films, since additional oxygen loss due to annealing in vacuum should intensify the observed trend instead of reversing it. The anneal processes were conducted in parallel with an oxygen deficient LSMO control sample to better understand oxygen loss process [37,38]. The resistance increases slightly after the 350 °C anneal in the unimplanted, oxygen deficient LSMO film and more substantially after the 400 °C anneal(Fig. 3c). The transition temperature and the high temperature insulating behavior are not affected. In contrast, the helium implanted film shows a consistent reduction in resistance and increase in transition temperature, even in the temperature regimes where oxygen is lost. A final anneal at 1 atm flowing oxygen at 650 ºC returns the implanted film to the undoped, as-grown character; however, this process does not affect the transition temperature in the oxygen deficient LSMO film. These results conclusively demonstrate that oxygen deficiency is not responsible for the decreased transition temperature and *c*-axis expansion in the helium implanted films and rule out the possibility that local crystal defects or amorphization is driving the observed resistive and magnetic changes, since the final anneal temperature is far below that necessary to recrystallize the LSMO film [17,39]. Most importantly, these results show that the implantation process is reversible so that very specific strain states can be controlled post-implantation and that helium is stable within the lattice far above room temperature.

Having ruled out oxygen non-stoichiometry, the observed changes in transition temperature could possibly be explained by disorder induced orbital ordering effects reported in bulk manganites [40,41]. Here, materials with various A-site cation compositions but with a common mean A-site radius show lower transition temperatures as the A-site variance is increased. However, the large changes in transition temperature observed in materials of high A-site variance in those systems are associated with an orthorhombic (O") to orthorhombic (O') structural phase change as discussed in references [40,41]. An increased variance in the orbitally disordered O" phase leads to a slight decrease in transition temperature of ~20°C. Once the variance surpasses a critical value, the orbitally ordered O' phase occurs and sharply drops the transition temperature by ~100°C while increasing resistivity by orders of magnitude. Increasing variance beyond this phase transition has only a weak effect on resistive behavior. Thus, variance induced disorder effects can be recognized by the sudden onset of structural reorientation producing large discontinous changes in transition temperature and resistive behavior. Moreover, magnetic moments tend to strongly decrease with increased variance [40,41]. The disorder mechanism is thusly ruled out for our results, since there is no evidence of a sudden changes in crystal structure, transition temperature, resistivity, or low temperature moment as a function of He implantation dose.

To understand how c-axis changes physical properties, we employ a Monte Carlo model to perform unbiased simulations and calculate the temperature-dependent resistivity using the Kubo formalism [17,22–24]. The LSMO film can be simulated using a two-orbital double-exchange model on a two-dimensional lattice, as previously confirmed to be appropriate for perovskite manganites and LSMO in particular [20,42]. The expansion along the *c*-axis is modeled by adjusting only a single term, the Jahn-Teller $Q_3$ mode, in the full Hamiltonian. It should be noted that model simulations on other systems involving biaxial strain and/or cation substitution require three or more terms to be modified [24].

To describe LSMO manganite films on $SrTiO_3$, a set of model parameters are adopted according to literature [23,43]. We fix the in-plane parameters for all thin films to be consistent with the epitaxial lattice locking in the experiment. The *c*-axis lattice expansion arising from the implantation of helium atoms is modeled through the Jahn-Teller distortion of the $Q_3$ mode, quantified as ~(*c*/*a*-1), which is the only term directly tied to the *c*-axis length. This simple shift of distortion lifts the energy degeneracy ($\Delta$) between the $x^2$-$y^2$ and $3z^2$-$r^2$ orbitals [23,24] thereby driving preferential orbital occupancy. Figure 4a shows the calculated orbital occupation of the $e_g$ electron for LSMO films across a range of $\Delta$'s. For high $\Delta$ values, there is a strong preference for the $x^2$-$y^2$ state, which decreases as $\Delta$ goes to zero. Figure 4b illustrates the orbital occupancy in a tensile strained film before and after *c*-axis elongation where the orbital lobe size denotes the relative preference of orbital occupation. An as-grown film has the largest $\Delta$ value due to the substrate-induced tensile strain which results in the largest energy difference between $x^2$-$y^2$ and $3z^2$-$r^2$ levels, where $\Delta = 180$ meV is consistent with previous experimental values given for epitaxial LSMO films on STO [26]. Increasing the out-of-plane expansion through He implantation while keeping the in-plane parameters constant decreases this energy gap and ultimately closes it in the cubic form where $\Delta = 0$. The Monte Carlo simulations show that with decreasing $\Delta$, i.e. with increasing *c*-axis lattice length, the resistivity is enhanced and the high temperature phase transition shifts to lower temperatures(Fig. 4c). For the unexpanded film case of $\Delta = 180$ meV, the system is metallic over the whole temperature region as expected. In the cubic $c/a = 1$ limit($\Delta = 0$), a prominent metal-insulator transition with high values of resistivity at high and moderate temperature regions is observed. These results are strikingly similar to the experimentally observed behavior. This suggests that He implantation may allow for an easy, highly tunable post-growth method of controlling orbital populations which could have an immediate impact on many fundamental studies of strongly correlated materials.

Thus, while it is possible to mimic the absolute volume and c/a ratio of the expanded LSMO samples using a combination of conventional lattice control techniques, ie. changing the unit cell volume through isovalent A-site substitution and control over the c/a ratio by epitaxial strain [44], the necessary modification of multiple parameters arising from the changes to the A-site pressure and Poisson effect contributions give rise to vastly different properties than those observed when only a single axis is manipulated without changing other parameters.

In summary, helium implantation offers a viable means to independently control the out-of-plane lattice parameter in a complex oxide film without directly altering other degrees of freedom. We find that this method of strain doping allows for very fine and continuous control over resistive and magnetic properties. Monte Carlo simulations of these unique lattice geometries indicate that the observed behaviors are driven by tuning the orbital occupation through Jahn-Teller effects. We emphasize that controllably tuning only a single lattice parameter will allow for a uniquely transparent evaluation of theoretical models across a broad range of materials by eliminating uncertainties inherent to the simultaneous manipulation of multiple degrees of freedom. Moreover, a critical step in bringing complex materials toward commercial applications is the ability to tune material properties using wafer-scale processing similar to current semiconductor technologies. The technique presented here demonstrates a path to achieving this need using strain doping, as it can be implemented using established ion implantation infrastructure in the semiconductor industry.

## Acknowledgements

This effort was supported by the US Department of Energy (DOE), Office of Basic Energy Sciences (BES), Materials Sciences and Engineering Division, (T.Z.W., C.B., W.S., J.D.B., A.H., P.C.S. and E.D.) and under US DOE grant DE-SC0002136 (A.W., H.W.G.). Helium implant preparations (P.D.R.) and magnetization measurements (Z.G.) were conducted at the Center for Nanophase Materials Sciences, which is sponsored at Oak Ridge National Laboratory by the Scientific User Facilities Division, Office of Basic Energy Sciences. Partial support for theoretical calculations was given by the National Science Foundation of China No. 11274060

(S.D.). R.T. and C.G.M acknowledge partial funding from the Joint Institute of Advanced Materials (JIAM), at the University of Tennessee.

**\*** To whom correspondence should be addressed: wardtz@ornl.gov

**Figure Legends**

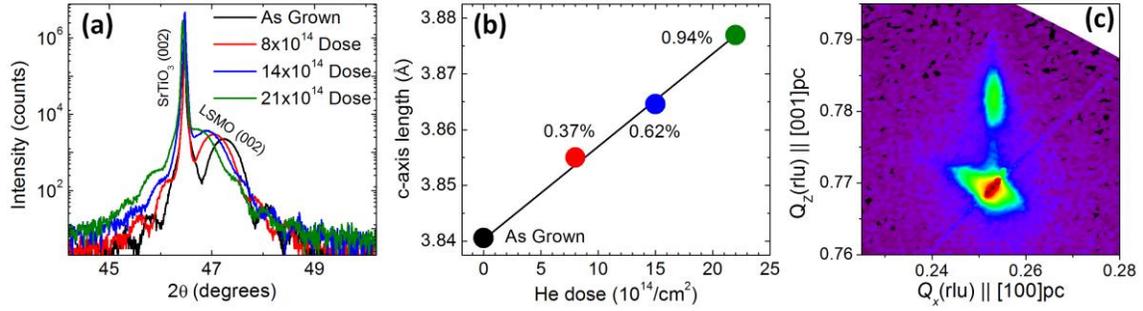

**Figure 1.** X-ray diffraction data on helium implanted LSMO thin films. (a)θ-2θ scans of LSMO thin films on STO substrates under different helium dosage given in ions/cm$^2$. (b)The out-of-plane lattice constant changes as a function of helium dose where percentages note the change in the *c*-axis relative to the undosed state. (c)Typical reciprocal-space mapping scan around the $(103)_{pc}$ peak of the $21 \times 10^{14}$ helium dosed sample shows that it remains epitaxially locked to the STO substrate.

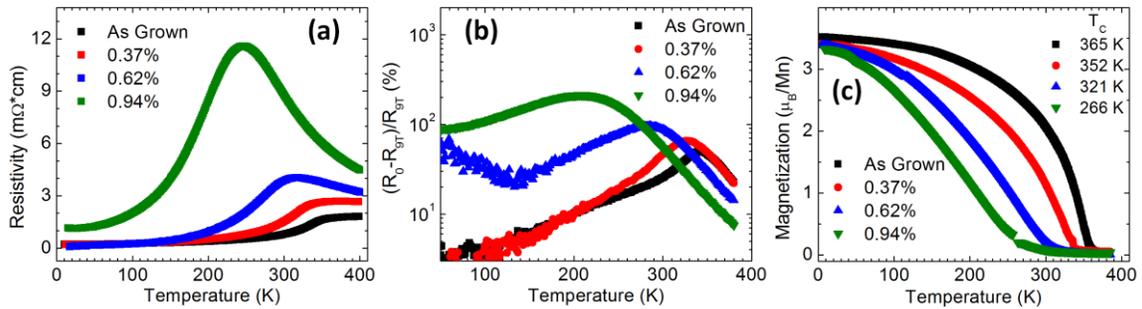

**Figure 2.** Transport and magnetization data of the LSMO thin films for different *c*-axis expansions. (a)Resistivity versus temperature measurements showing a reduced phase transition temperature and enhanced resistance as *c*-axis expands. (b)Maximum magnetoresistance values increase as the *c*-axis expands. (c)Corresponding magnetization measurements also show a decrease in $T_C$ with increasing *c*-axis expansion.

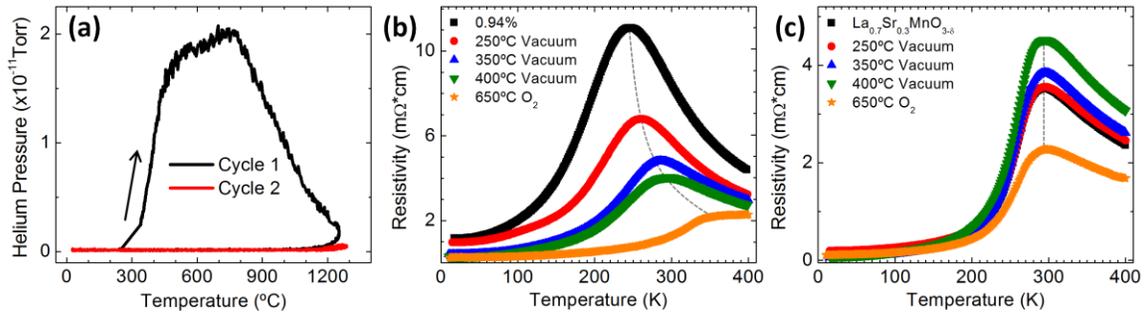

**Figure 3.** Results of annealing He implanted LSMO and oxygen deficient LSMO films. (a)Residual Gas Analysis of an implanted film under consecutive heating/cooling cycles shows that He is released from the film above ~250 ºC. (b)Resistive behavior of the 0.94% LSMO sample after vacuum annealing. A 650 ºC anneal under 1atm flowing oxygen is sufficient to return film to nearly identical character to the as-grown state [17]. (c)An oxygen deficient LSMO film under these same annealing conditions for comparison. Dotted lines are drawn connecting $T_{MIT}$, showing a very clear shift toward the unimplanted state in the 0.94% sample as He is evacuated, while $T_{MIT}$ is unaffected in the oxygen deficient film.

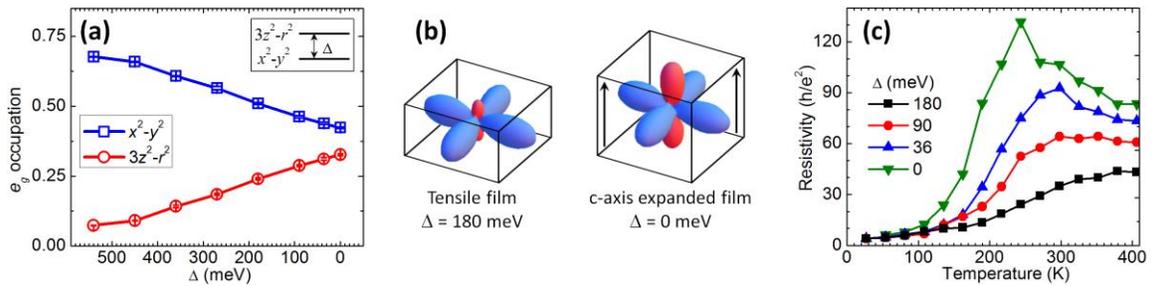

**Figure 4.** Monte Carlo simulation of the LSMO double-exchange model. (a)Relative orbital occupancy as a function of the energy difference($\Delta$) between $x^2$-$y^2$ and $3z^2$-$r^2$ orbitals. The orbital occupancy of the $e_g$ orbitals is shifted from $x^2$-$y^2$ in a tensile strained epitaxial film to $3z^2$-$r^2$ by increasing $c$-axis expansion(decreasing $\Delta$). (b)Diagram of orbital occupancy where size of lobes indicates relative filling. A tensile strained film has a preference toward the $x^2$-$y^2$ orbital (blue lobes) over the $3z^2$-$r^2$ orbital (red lobes). (c)Resistivity versus temperature for strained films with increasing out-of-plane lattice expansions that are given in terms of the induced Jahn-Teller splitting between the two $e_g$ orbitals($\Delta$) . With decreasing $\Delta$, i.e. with elongating $c$-axis, the phase transition temperature is reduced and resistivity is enhanced.